\documentclass[aps,prl,preprint,groupedaddress,showkeys,floatfix]{revtex4-1}

\usepackage[english]{babel}
\usepackage[utf8x]{inputenc}
\usepackage[T1]{fontenc}

\usepackage{siunitx}
\usepackage{amsmath}
\usepackage{graphicx}
\usepackage[colorinlistoftodos]{todonotes}
\usepackage[colorlinks=true, allcolors=blue]{hyperref}

\begin{document}

\title{Microrheology of DNA hydrogels}

\author{Zhongyang Xing}\email{zx230@cam.ac.uk}
\affiliation{Optoelectronics, Cavendish Laboratory, University of Cambridge, Cambridge CB3 0HE, United Kingdom}
\author{Alessio Caciagli} \thanks{These authors contributed equally to this work}
\affiliation{Optoelectronics, Cavendish Laboratory, University of Cambridge, Cambridge CB3 0HE, United Kingdom}
\author{Tianyang Cao} \thanks{These authors contributed equally to this work}
\affiliation{Department of Chemistry, University of Tsinghua, Beijing, 100084, China}
\author{Iliya Stoev}
\affiliation{Optoelectronics, Cavendish Laboratory, University of Cambridge, Cambridge CB3 0HE, United Kingdom}
\author{Mykolas Zupkauskas} 
\affiliation{Optoelectronics, Cavendish Laboratory, University of Cambridge, Cambridge CB3 0HE, United Kingdom}
\author{Thomas O'Neill}
\affiliation{Optoelectronics, Cavendish Laboratory, University of Cambridge, Cambridge CB3 0HE, United Kingdom}
\author{Tobias Wenzel}
\affiliation{Optoelectronics, Cavendish Laboratory, University of Cambridge, Cambridge CB3 0HE, United Kingdom}
\author{Robin Lamboll}
\affiliation{Optoelectronics, Cavendish Laboratory, University of Cambridge, Cambridge CB3 0HE, United Kingdom}
\author{Dongsheng Liu}
\affiliation{Department of Chemistry, University of Tsinghua, Beijing, 100084, China}
\author{Erika Eiser}\email{ee247@cam.ac.uk}
\affiliation{Optoelectronics, Cavendish Laboratory, University of Cambridge, Cambridge CB3 0HE, United Kingdom}

\begin{abstract}
	A key objective in DNA-based material science is understanding and precisely controlling the mechanical properties of DNA hydrogels. We perform microrheology measurements using diffusing-wave spectroscopy (DWS) to investigate the viscoelastic behavior of a hydrogel made of Y-shaped  DNA nano-stars over a wide range of frequencies and temperatures. Results show a clear liquid-to-equilibrium-gel transition as the temperature cycles up and down across the melting-temperature region for which the Y-DNA bind to each other. These first measurements reveal the crossover of the elastic $G'(\omega)$ and loss modulus  $G''(\omega)$ when the DNA-hydrogel formed at low temperatures is heated to a fluid phase of DNA nano-stars well above the melt temperature $T_m$. We show that the crossover relates to the life-time of the DNA-bond and also that percolation coincides with the systems' $T_m$ . The approach demonstrated here can be easily extended to more complicated DNA hydrogel systems and provides guidance for the future design of such transient, semi-flexible networks that can be adapted to the application of molecular sensing and controlled release. 
\end{abstract}

\date{\today}  % the date we submit

\keywords{DNA hydrogels, self-aseembly, microrheology, diffusing wave spectroscopy, trasient polymer network}

\maketitle  % make the title

\section{\label{suc:introduction}Introduction}
DNA hydrogels are a type of tenuous, semi-flexible polymeric network that consists of precisely designed synthetic nucleotide strands as chemical or physical cross-linkers \cite{luo2006enzyme-dna-gel,luo2012mechanical-dna-gel, liu2017dna-programmable, liu2014dna-i-motif-reveiw, liu2011self}. These man-made bulk DNA hydrogels have been widely studied as functional materials that can be potentially used for controlled drug delivery, tissue engineering, biosensing and other applications in the fields of nanotechnology and bioengineering mainly becasue of their bio-compatibility and the ability to mix them with other (bio)polymers \cite{okay2011dna,liu2015dna_bio-printing,liu-oren2015dna-CB8,liu2015writable-polypeptide-dna,liu2013dna-release-cells}. In particular, the vast combinations of the Watson-Crick pairing provide a unique way to achieve programmable self-assembly of thermally reversible gels with precise functionality \cite{rovigatti2014gels,liu2017dna-hg-review}. Current studies mostly focus on the fabrication and utilization of DNA hydrogels, while the fundamental physics of these gels still lacks good understanding \cite{liu2011self,liu2016reversibly,liu-oren2015dna-CB8}. In recent years, a series of computational and experimental studies were carried out on the phase diagram of DNA hydrogels made of 2-, 3- and n-valent nano-stars and their mixtures, which provides a good reference for creating volume-spanning, percolating gels \cite{rovigatti2014gels,biffi2013phase,biffi2015equilibrium,rovigatti2014accurate,fernandez2016small,nava2017fluctuating}. 

The study of transient networks has been at the heart of many theoretical\cite{curro1983theoretical,groot1995dynamic,edwards1987gels} and experimental studies \cite{filali2001robust,porte2006bridging} for their display of complex phase diagrams and dynamics\cite{koike1995dynamic}. Different to chemically crosslinked networks such as rubbers, the crosslinks in transient networks can be mediated by the short sticky ends of telechelic polymers \cite{michel2000percolation}, telechelic dendrimers\cite{likos2009telechelic}, triblock-copolymers\cite{gabor2004telechelic} or charged end-groups\cite{eiser1999shear}, which have a finite life time, thus rendering these networks yield-stress fluids. Other transient but active networks are formed by the semiflexible actin filaments crosslinked via proteins are of great importance for giving shape to cells and their locomotion\cite{vahabi2017normal}. Vitrimers, in which the bonds or crosslinks can be exchanged through a catalytic process, are another class of transient networks with self-healing properties relevant in biological tissue engineering\cite{Leibler2011silica,Laibler2012catalytic,Leibler2014nanoparticle}. With their small building blocks, DNA nano-stars present very similar water-based networks, however, with much greater versatility. Recent rheological studies\cite{pan2016effects,li2010slow,liu2017dna-programmable} in the low frequency range showed that the specific structure and connectivity of the DNA-nanostars has a strong influence on their macroscopic mechanical response. However, a detailed understanding of the relation between a repeat-unit and the overall macroscopic response and the related relaxation processes is not yet fully understood. Here we present micro-rheology studies that shed light on these aspects, which aim to be of great importance in the design of new functional materials\cite{luo2012mechanical-dna-gel}.

Diffusing-wave spectroscopy (DWS), a technique that measures the dynamics of strongly scattering samples\cite{pine1988dws,mason1995optical} is an established technique to perform microrheological studies of complex fluids at much shorter time and length scales than conventional bulk-rheology can achieve \cite{mason1997osa,waigh2005microrheology-review,squires2010fluid-review,galvan2008diffusing}. Here we present DWS microrheology studies of DNA hydrogels, probing the rheological properties in a dramatically larger frequency range than the few studies done so far using bulk rheology, where only the effect of built-in flexibility in the joints was measured at the long-time response of the gel \cite{pan2016effects,niu2016langmuir}. Typical short-time relaxations inherent to the fast making and breaking times of the transient DNA-gel in the melt-transition region can not be accessed in bulk rheology. We elucidate the relation between the system's characteristic binding-unbinding processes and the local and global mechanical properties of the gel over a temperature range that covers the full melting region between the DNA nanostars. 

In this paper, we first introduce the design principle of the Y-DNA building blocks that form the hydrogel and their characterization. Secondly, we present DWS measurements of our Y-DNA  system in a concentration range showing a continuous transition from a fluid to equilibrium-gel phase, presenting $G'(\omega)$ and $G''(\omega)$ over a broad range of frequencies and temperatures. We choose this system because it represents an efficient model to investigate the general physics of an n-valent DNA hydrogel with varied rigidities, percolation behaviour and valency. These results will provide strong guidance to future design of more complicated DNA hydrogel frameworks with controlled mechanical properties.

\section{\label{sec:results}Results}

\begin{figure*}[tbhp]
	\centering
	\includegraphics[width=1\textwidth]{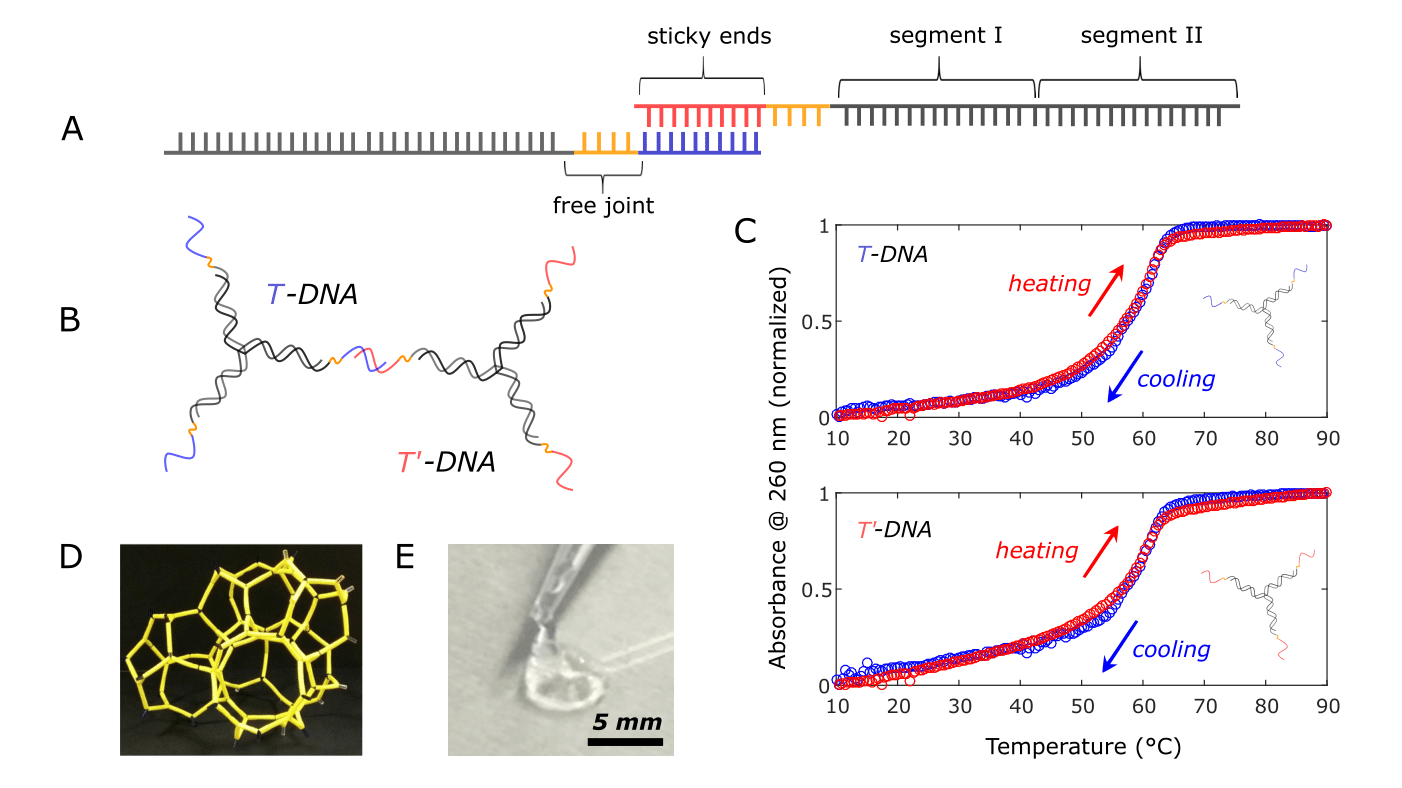}
	\caption{Design and characterization of the DNA hydrogel building blocks. (A) Schematic of the ssDNA \textbf{S}$_i$ used. Each oligo strand consists of four functional parts: the \textit{sticky end}, \textit{free joint}, \textit{segment I} and \textit{segment II}. \textit{Segment I \& II} are part of the double stranded core-DNA; \textit{sticky ends} are for cross-linking the Y-shapes to a network. (B) Cartoon of \textbf{T} and \textbf{T'} DNA connected via hybridization of complementary \textit{sticky ends}. (C) Melting (cooling) and heating (hybridization) curves of \textbf{T} and \textbf{T'} DNA in Tris-EDTA buffer containing 150 mM NaCl, measured using by UV-vis spectroscopy. (D) A macroscopic model assembled by laser-cut Y-shaped units and connected through elastic, yellow tubes illustrating a possible network formation. (E) Snapshot of a DNA hydrogels made of [\textbf{T} DNA]=[\textbf{T'} DNA]=500 \si{\micro M}. The sample maintains its original shape at a time scale of several seconds.}
	\label{fig:dna-gel}
\end{figure*}

\subsection{\label{subsec:dna-hydrogels}Design of DNA hydrogels}
The DNA hydrogels used in this study are composed of Y-shaped DNA building blocks made of three partially complementary oligonucleotides, here denoted  \textbf{S$_i$} (Fig. \ref{fig:dna-gel} and Table \ref{table:DNA-sequences}). These single-stranded S-oligos are 41 bases long and consist of four functional parts: a \textit{sticky end} (7 bases), a \textit{free joint} of 4 Thymines providing flexibility, and two segments (I \& II) that form the core of the Y-shape.  Segments I and II of say ssDNA \textbf{S}$_1$ are designed to be partially complementary to the respective segments of \textbf{S}$_2$ and \textbf{S}$_3$, thus making up the three dsDNA arms at the centre of the Y-shape. Note that the center is created to be fully binding, leaving no non-binding bases at the center thus reducing its flexibility. The respective binding sequences of the three dsDNA arms are given in Table  \ref{table:DNA-sequences} while the full sequences of the \textbf{S}-oligos are given in the SI. The \textit{sticky ends} can specifically bind to their complementary sequence due to Watson-Crick pairing. In all following gels we use the same Y-shape cores carrying either \textit{sticky ends}  named \textbf{T} or the complementary \textbf{T'} DNA, such that Y-shapes with \textbf{T}-ends can only bind to those with \textbf{T'}-ends.  We keep the \textbf{T} to  \textbf{T'} DNA ratio 1:1, thus maximizing the system's connectivity. 

The melting temperature profiles of the Y-shapes carrying either \textbf{T} or \textbf{T'} sticky ends (Fig. \ref{fig:dna-gel} C) were determined by measuring the absorbance of 260 \si{nm} in 1 \si{\micro M} DNA in Tris-EDTA buffer solution at pH = 8.0 containing 150 mM NaCl, as ssDNA adsorbs 260 \si{nm} stronger than dsDNA. Starting at 90 \si{\celsius} we observe a plateau in the absorbance until roughly 65 \si{\celsius}, marking the point at which hybridization (binding) sets in. Until this temperature the individual ssDNAs are all in the unbound state.  Upon further cooling the absorption decreases continuously until the low temperature plateau is reached at which all single strands have hybridized into Y-shapes. The melt temperature (denoted $T_{m1}$) is defined as the point at which half of all possible base pairs are dissociated. $T_{m1}$ was obtained from the median between the linear curves fitting the low- and high-temperature plateaus and found to be $T_m \approx$ 58 \si{\celsius} for both Y-shapes with \textbf{T} or \textbf{T'} overhangs, for the concentrations used in the melt temperature measurements. As shown previously \cite{Lorenzo-2014-JACS},  non-binding ssDNA tails slightly lowered the value of $T_m$ with respect to the value based on tabulated data by Santalucia, which is, averaged over all three arms, $T_{m1} \geq$ 60 \si{\celsius} \cite{santalucia1998unified} - details are given in the SI. 

\begin{table}[h!]
 \caption{The sequences of dsDNA arms}
 \label{table:DNA-sequences}
 \centering
 \begin{tabular} {c|c c c}
 \hline
 name &segment I & segment II \\
 \hline
 \textbf{S}$_1$&5'-TGG ATC CGC ATG ATC&CAT TCG CCG TAA GTA-3'\\
 \textbf{S}$_2$&5'-TAC TTA CGG CGA ATG&ACA CCG AAT CAG CCT-3'\\
 \textbf{S}$_3$&5'-AGG CTG ATT CGG TGT&GAT CAT GCG GAT CCA-3’\\
 \hline
 \end{tabular}
%\addtabletext{}
\end{table}

\subsection{DWS Microrhelogy}
\label{subsec:dws}

DWS measurements rely on time correlations of diffusively scattered light caused by submicron-sized spherical tracer particles, here 600 nm large, sterically stabilized polystyrene (PS) particles, embedded in the DNA hydrogel sample. The thermal motion of the particles is described by the mean-square displacement (MSD), denoted as $\left\langle\Delta r(t)^2\right\rangle$. This is measured by the real-time fluctuations of the scattered light collected by a photodetector and presented as the intensity-autocorrelation  function (ICF), $g_2(\tau) = \left\langle I(\tau)I(0)\right\rangle_t/\left\langle I(0)\right\rangle^2_t$, where $I(0)$  and $I(\tau)$ are the scattered intensities at time zero and some delay time $\tau$ later. The decay of $g_2(\tau)$ is related to the time evolution of particle motion, allowing for the MSD of the particle to be measured \cite{pine1988dws}. The complex viscoelastic modulus and MSD are related by the generalized Stokes-Einstein relation (GSER) given in eqn.\ref{eqn:GSER} based on the fluctuation-dissipation theorem \cite{mason1997osa,mason1995optical}, 
\begin{equation}
	\tilde{G}(s)\approx\frac{k_BT}{\pi Rs\left\langle \Delta \tilde{r}^2(s)\right\rangle}, 
	\label{eqn:GSER}  
\end{equation}
where \textit{\~G(s)} and $\left\langle\Delta \tilde{r}(s)^2\right\rangle$ are the Laplace transforms of the complex shear modulus and MSD; \textit{R} is the radius of the tracer particles, and \textit{s} the Laplace frequency. Replacing $s$ by $i\omega$ we obtain the $G'(\omega)$ and $G''(\omega)$ from the complex shear modulus $G^*(\omega)=G'(\omega)+iG''(\omega)$ \cite{mason1995optical,mason1997osa}.

\begin{figure}[h!]
	\centering
	\includegraphics[width=0.8\linewidth]{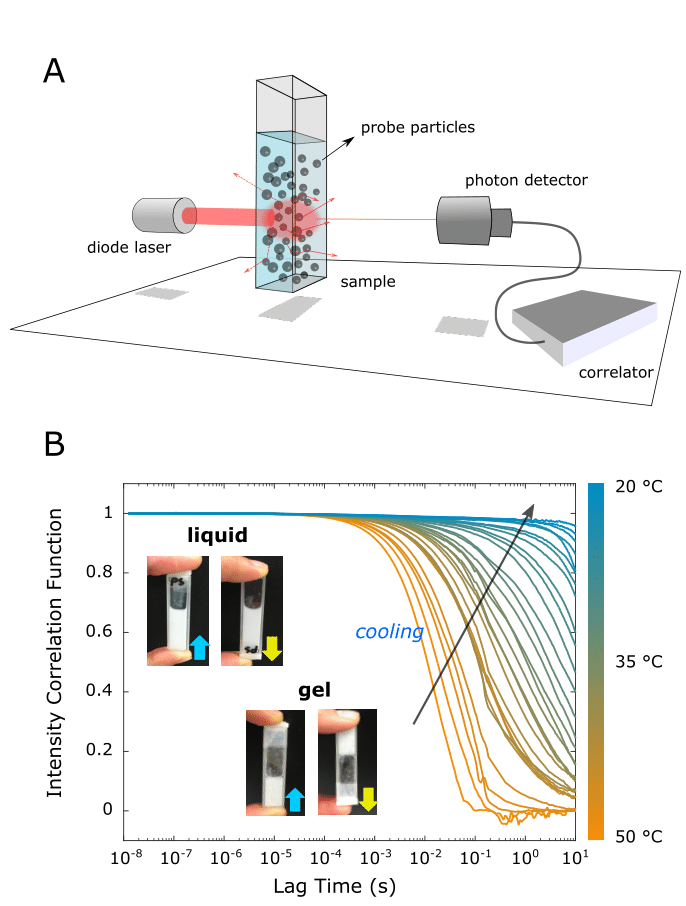}
	\caption{(A) Schematic illustration of the diffusing-wave spectroscopy (DWS) setup: A 685 nm diode laser beam impinges onto the sample, and the diffusely scattered light is collected by the photodiode on the other side of the sample. Tracer particles are uniformly embedded inside the sample. (B) Temperature-dependent intensity-correlation function (ICF) curves measured for the 500 \si{\micro M} DNA hydrogel, containing 1 v/v\si{\percent} 600 nm large, sterically stablised polystyrene tracer particles. The ICF curves were measured starting from 50 \si{\celsius} (orange lines) in 1 \si{\celsius} steps cooling down to 20 \si{\celsius} (blue lines). The photographs show the sample cuvette showing the samples liquid state at 50 \si{\celsius} \textit{top left} and the gel state at 20 \si{\celsius} to \textit{bottom rightt} .}
	\label{fig:icf}
\end{figure}

Our DNA hydrogels were studied using DWS in echo-mode of the Light Instrument Reheolab allowing for long correlation times (Fig. \ref{fig:icf} A). The light source was a  685 nm wavelength diode laser. The measured ICF where fitted and then converted into a mean-squared displacement (MSD) using both the instrument software and for comparison a home-written software. The procedures are detailed in the SI. 

DNA-hydrogel samples were made by preparing \textbf{T}- and  \textbf{T'}-DNA solutions separately both at a DNA concentration of 500 \si{\micro M} (see Methods), already containing the tracer particles. The cuvettes (Fig. \ref{fig:icf} B) were filled with the respective \textbf{T}- and  \textbf{T'}-DNA solutions, which are liquid at room temperature (RT) in a layer-by-layer fashion at RT for better mixing. The total filling volume was $\sim$ 500 \si{\micro l}. Initially the interfaces between the layers gelled due to the rapid hybridization at RT. To obtain well mixed samples we heated the cuvettes to 50 \si{\celsius} and incubated them for 20 minutes, which was sufficient to fully melt the sticky ends rendering the sample a well-mixed fluid of Y-shapes, but also cold enough to preserve the Y-shaped structure. Following this procedure we ensured that the final DNA concentration in the fluid phase remained 500 \si{\micro M}, corresponding to $\sim$20 mg/ml, containing 1 \si{\percent} w/v PS particles.

DWS measurements were  done starting at 50 \si{\celsius} and slowly cooled down to 20 \si{\celsius}. We measured the scattering intensity in 1 \si{\celsius} intervals, and equilibrate the sample for 5 min prior to data acquisition, ensuring the full coverage of the melting transition of the sticky ends. The scattered intensities were then converted into ICFs and MSDs. We performed control experiments using 230 nm large PS particles testing both for a possible dependence of our results on particle size and for hysteresis in the gelation process by repeating the scattering measurements in a cooling and heating cycle following the same protocol as for the larger particles. The results, shown in the SI, confirm that we only probe the viscoelastic properties of our DNA-hydrogel and that it is a thermally reversible equilibrium gel. The ICF data for the cooling ramp are shown in Fig. \ref{fig:icf} B.

\section{Discussion}
\label{discussion}

\begin{figure}[h!]
	\centering
	\includegraphics[width=.618\linewidth]{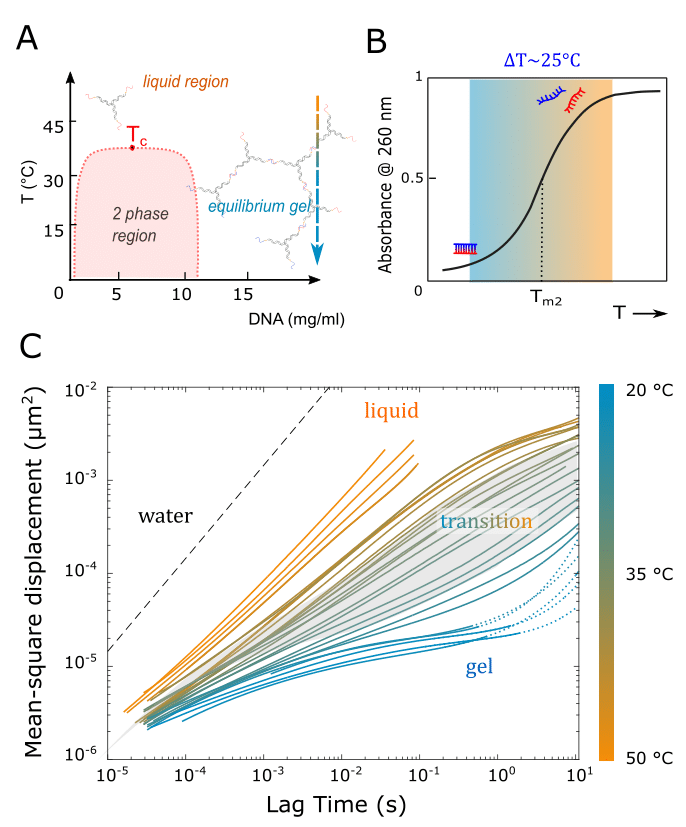}
	\caption{(A) Schematic phase diagram of the Y-shaped DNA. The arrow indicates the concentration and temperature range of the ICF measurements shown in Fig. \ref{fig:icf}. The red area represents the two-phase region, and $T_c$ the critical point. (B) Illustration of the hybridization range for the sticky ends of the two different Y-shapes. The graded area signifies the range over which a fraction of base-pairs are formed. (C) Mean-square displacement (MSD) extracted from the ICF curves in Fig. \ref{fig:icf}. The color of the lines gradually changes from orange to blue, standing for the transition region ranging from about 45 \si{\celsius} to 20 \si{\celsius}, which is centered around the melting temperature $T_{m2}$ = 35 \si{\celsius} of the sticky ends. The calculated MSD for the same 600 nm large PS colloids in pure water at 50 \si{\celsius} is presented by the dashed line as guide to the eye.}
	\label{fig:msd}
\end{figure}

Several simulation and experimental studies explored the phase diagram of DNA-hydrogels made of nano-stars with 2-, 3- and more bridging arms \cite{biffi2013phase,biffi2015equilibrium,rovigatti2014accurate,locatelli2017condensation}. These studies showed that the region where a phase-separation into a nano-star poor and nano-star rich region becomes increasingly more narrow, shifting to the lower DNA-concentration end for decreasing valencies \cite{biffi2013phase}. Here we used a total DNA concentration of 500 \si{\micro M} ensuring that we are firmly in the one-phase region (Fig. \ref{fig:msd} A) with the hydrogel spanning the entire sample. This entails that as we cool the sample from 50 \si{\celsius} to RT the entire sample is brought continuously from a fluid phase containing free Y-shapes to a percolating gel network. In order to obtain a clear signature of this network formation its melt temperature $T_{m2} \approx$ 35 \si{\celsius} must be well below $T_{m1}$ of the individual Y-shapes. This melt-temperature separation was achieved by increasing the added salt concentration to 200 mM NaCl \cite{Lorenzo-2014-JACS}. The melting temperature will also depend weakly on the DNA concentration shifting our systems'  $T_{m1} $ to $\sim$ 62 \si{\celsius}, but cannot be measured directly as such high DNA concentrations are adsorbing too strongly. Also $T_{m2}$ cannot be measured by UV adsorption as its signal is small compared to that of the dsDNA in the Y-shape. More details of the $T_{m2}$ estimate are given in the SI.

The MSD results extracted from the ICF curves are shown in Fig. \ref{fig:msd} C. At temperatures well above $T_{m2}$  (orange lines), the MSD curves depend linearly on the lag time $\tau$ over the whole measured region, confirming the $\left\langle\Delta r(t)^2\right\rangle \propto \tau$ relation for Newtonian fluids and thus proving that in the temperature window between about 55 \si{\celsius} and 45 \si{\celsius} our system behaves like a fluid of disconnected Y-shapes dispersed in a buffer-colloid solution. However, compared with the calculated MDS for 600 nm PS particles dispersed in pure water with a diffusion constant $D = 0.15$ \si{\square\micro \meter\per\second} at 50 \si{\celsius} our sample displays a diffusivity that is two orders of magnitude lower. A similar decrease is observed in the control measurements using 230-nm large PS particles (see SI). This decrease cannot be due to the presence of the colloids, which would lower the diffusion coefficient only by 2\si{\percent} according to the Einstein relation for the viscosity of colloidal dispersions or the buffer conditions. Hence the observed increase in viscosity is purely due to the high DNA-concentration. Indeed, assuming that the Y-shapes take up an effective spherical volume due to rotational diffusion (assuming that each arm is $\sim$ 5 nm long) the approximate volume fraction occupied by the Y-shapes is some 40\%, although the actual DNA content is only 2 weight\%.

\begin{figure*}[ht]
	\centering
	\includegraphics[width=1\textwidth]{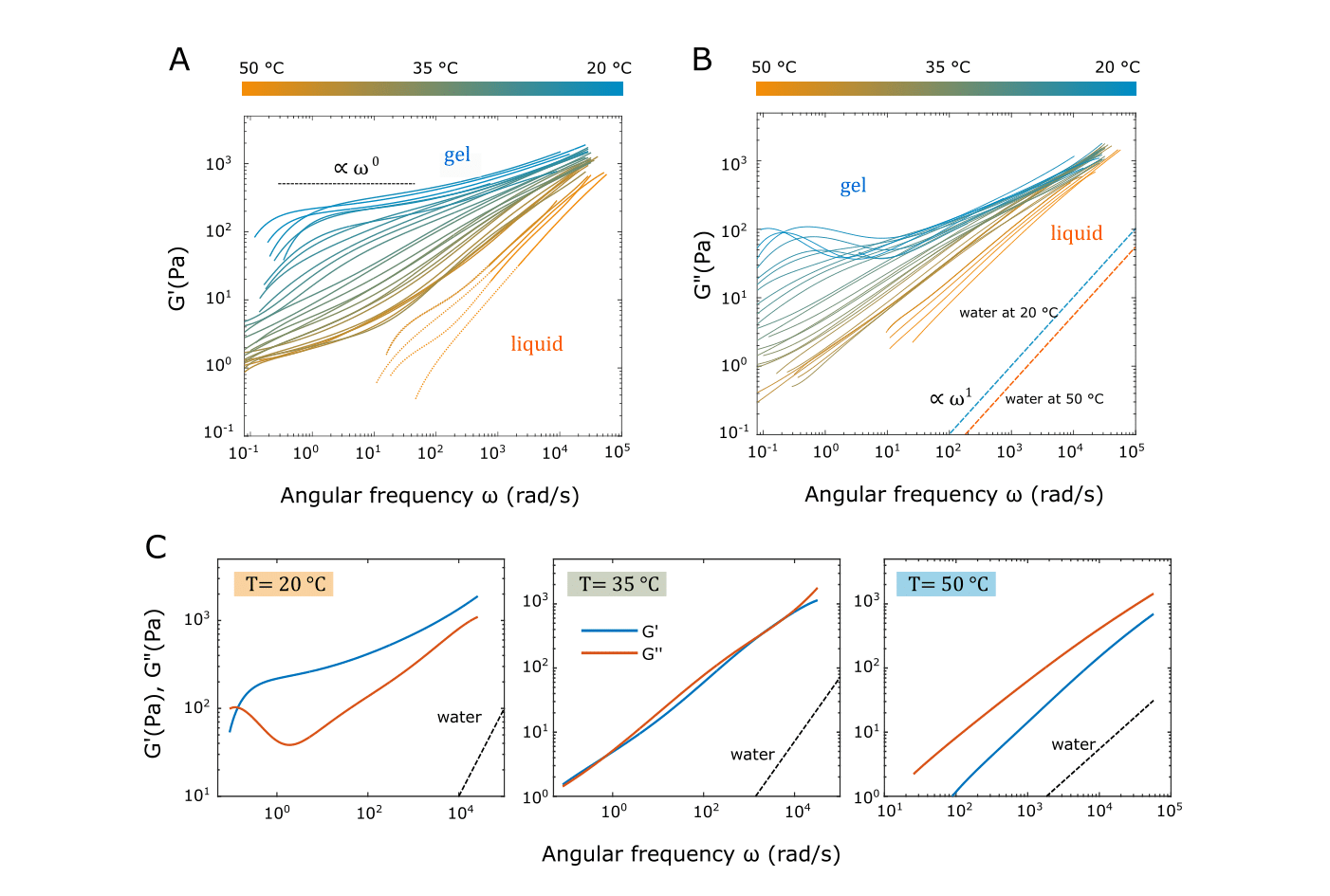}
	\caption{Temperature-evolution of the complex moduli $G'(\omega)$ and $G''(\omega)$ as a function of frequency extracted from the MSDs in Fig. \ref{fig:msd}. (A) The elastic moduli $G'(\omega)$ measured in a cooling ramp. At temperatures above $T_{m2}$, $G'$ drops down at a frequency below $10^2 \sim 10^4 rad/s$, showing close-to-zero elasticity; below $T_{m2}$, only th eonset of the decay in the ICF could be monitored. Hence the low-frequency region is plotted as dashed lines using extrapolations. (B) The viscous modulus $G''(\omega)$ in cooling (\textit{top}) ramp. (C) Comparison of $G'(\omega)$ and $G''(\omega)$ at temperatures of 20 \si{\celsius}, 35 \si{\celsius}, and 50 \si{\celsius}, representing typical behavior at temperatures below, around, and above $T_m$. At 20 \si{\celsius}, $G''$ is higher than $G'$ at frequencies below the crossover frequency $\sim 10^4$ rad/s; at 35 \si{\celsius}, $G'$ and $G''$ are overlapping over almost the entire frequency range; at 50 \si{\celsius}, $G''$ is higher than $G'$ over the whole measurable frequency range, showing no crossover point at all.}
	\label{fig:complex moduli}
\end{figure*}

At $T <$ 30 \si{\celsius} and short lag times, the MSD curves are similar to the high-$T$ measurements increasing with increasing $\omega$, however with a slightly lower exponent indicating subdiffusive motion of the local bridges between crosslinks. At intermediate $\omega$ corresponding to longer relaxation times, for instance of the 'cages' formed by the crosslinks, the MSD curves reach a plateau. Holding the sample at this lower temperature over 20 minutes and measuring the ICF in 5 minutes' intervals show that there is no further increase in the plateau value (see SI). This is also expressed in the flattening of the corresponding elastic moduli $G'$ presented in Fig. \ref{fig:complex moduli}. This means the tracer particles remain locally diffusive on short time scales (the diffusion coefficient of the particles in pure water is 1.65 \si{\square\micro\meter\per\second} at 20 \si{\celsius}) , but are confined by the percolating DNA-network on long time scales. The transition region marked by the changing colors in Fig. \ref{fig:msd} and Fig. \ref{fig:complex moduli} represents the melt-temperature region over which the fraction of hydrogen bonds formed between two Y-shapes with complementary sticky ends is gradually increasing as $T$ decreases. Using refined Santalucia rules for hybridization \cite{Lorenzo-2014-JACS} we estimate the width of this transition region to be $\Delta T \sim$  25 \si{\celsius} (see Fig. \ref{fig:msd} B). With a $T_{m2} \sim$ 35 \si{\celsius} this means we should reach a fully bonded state and thus a maximum network stiffness at around $T \sim$ 25 \si{\celsius}. A detailed discussion of the value of the stiffness is given in the following. 

The elastic, $G'(\omega)$, and viscous moduli, $G''(\omega)$, measured in a cooling cycle are shown in figure \ref{fig:complex moduli}. Again, similar results were obtained using the smaller tracer particles in a cooling and heating cycle (see SI), suggesting that the equilibrium gels display only very small hysteresis effects. As expected, the elastic modulus, $G'(\omega)$ undergoes a significant change as the temperature changes over the melting-temperature region, while $G''(\omega)$ retains the same linear trend until about 30 \si{\celsius}. At high temperatures (orange lines), $G'$ is nearly zero at long time scales as the solution is in a fully fluid state of Y-shapes, but is non-zero at high frequencies reflecting the fact that  the sample is a quasi-concentrated solution of elastic-shapes behaving like soft colloids. Indeed, above $T_{m2}$ the loss modulus dominates (Fig. \ref{fig:complex moduli} C). However, around the melting temperature (between 37-31 \si{\celsius}),  $G'(\omega)$ and $G''(\omega)$ run parallel and on top of each other, which we identify as the point of full percolation. This  percolation can be understood when looking at Fig. \ref{fig:msd} B: in the melt-transition region increasingly more Y-shapes bind to each other forming many clusters that grow in size as the temperature decreases. At $T_{m2}$ half of all possible hydrogen bonds  or base pairs are bound, which does not mean that half of all Y-shape arms are bound at all times but that they continuously form and break partially and thus on average form a single cluster. Below $T_{m2}$ the fraction of hybridized base pairs continues to increase until about 25 \si{\celsius} and also their life time becomes longer. At even lower temperatures (blue lines), the $G'$ reaches a plateau value of $\sim$ 400 Pa in the intermediate time range, which corresponds to a mesh size $\xi \sim$ 21.5 nm, assuming the scaling behaviour of the bulk modulus $G_{bulk} \propto k_BT/ \xi^3$. This is in good agreement with the calculated mesh size from the design that suggests an average distance between bonded Y-shape centres of $\xi \sim$ 20 nm corresponding to a slightly higher elastic modulus. 

Interestingly, below $T \sim$ 25 \si{\celsius} our fully formed network is very similar to that of classical transient networks of flexible polymers held together by crosslinks through physical interactions that constantly form and break (Fig. \ref{fig:complex moduli} C) \cite{groot1995dynamic}. The frequency behaviour of such transient networks shows a typical Maxwellian $G''(\omega) \propto \omega$ increase at long times, reflecting the fact that the \textbf{TT'} bonds between the Y-shaps are not irreversibly formed. In fact, we see a crossover between the steeper increasing $G'(\omega)$ and $G''(\omega)$ at around 1 \si{\per\second}, which is know to correspond to the life time of the short double-stranded bond between two Y-shapes at RT, and can be explained by the reptation model for associative polymers \cite{edwards1987gels,curro1983theoretical}. Note that we cannot determine the exact crossover as the errors in our DWS measurements start to diverge at smaller frequencies. However, as expected this crossover frequency increases slightly as the temperature rises, reflecting the decrease in the bond life-time (Fig. \ref{fig:complex moduli} C). A similar observation was made in dynamic light scattering experiments on gels of 4-armed nanostars \cite{nava2017fluctuating}. This crossover is then followed by a plateau in $G'(\omega)$, while $G''(\omega)$ decreases slightly for some intermediate frequency range before both moduli start increasing again at higher frequencies. In our case this dip in $G''(\omega)$ can be ascribed to the fact that the elastic network dominates in this time regime before a sufficient number of bonds break and the tracer colloids can move. Consequently, this dip in $G''(\omega)$ disappears when the temperature reaches the melt-transition region when the life time of the bonds decreases further and the viscous contributions increase. At even higher frequencies $G''(\omega)$ gradually reaches the linear behaviour in $\omega$ as it reaches the fully liquid state by coming from a weakly sub-diffusive behaviour in the transition region, while $G'(\omega)$ reaches a power-law behaviour with a fractal power. In the gel phase this upturn in $G'(\omega)$ at high frequencies is simply due to the fact that the bond life-time is now longer than the typical motion due to thermal fluctuations and thus the system shows the typical increase in elasticity as the system cannot relax fast enough. This is associated with its second crossover between the two moduli appears to occur at $\omega_c \approx 4\times 10^4$ Hz at temperatures at $T \sim$ 35-50 \si{\celsius} (Fig. \ref{fig:msd} C). Estimating a second characteristic length scale $\xi' = (2k_BT/(\eta \omega_c))^{-3} \approx$ 34 nm, assuming the viscosity of water. At this high frequency the main elastic contribution must come from the cluster phase, with clusters forming `cages' of the Y-shapes. These completely disappear at even higher temperatures. 

Finally, at $T \gtrsim$ 35 \si{\celsius} we can plot the half-time of the relaxation of the ICF as inverse function of temperature. The slope of the resulting Arrhenius plot provides us with the strength of the bonds between two Y-shapes, where the relaxation time $\tau = t_{1/2} = \tau_0 \exp(-\Delta H/k_BT)$. Following the arguments by Nava et al. this will happen when at least 2 bonds per Y-shape are broken, which corresponds to about 60 kcal/mol in our case.

To summarise, our micro-rheological measurements demonstrate how a transient cross-linked hydrogel is formed as it is brought from the high temperature range, where the Y-shaped building-blocks form a viscous fluid, into an equilibrium gel-phase. Remarkably, once all possible bonds are formed below the melt-transition region the DNA-hydrogel shows a frequency behaviour very similar to that of transient networks of flexible polymers, although the connection between two Y-shape centres is not a fully flexible polymer but rather two semi-stiff dsDNA strands connected via two short, flexible ssDNA linker and another rigid dsDNA (\textbf{TT'}) bond. Interestingly, when the flexible ssDNA linkers are removed the elasticity of such a network, using otherwise the same DNA strands and concentration, increases to the point that we could not measure the inherent elasticity of the system. We also remark on the interesting transition-region, in particular around the melt-temperature $T_{m2}$, where we could identify the percolation transition. Finally, different regions in the measured frequency range could be associated with characteristic bond life-times, which to our knowledge is the first time. From RT to above the transition-region the bond's life time increases steadily until we hit the region where the system is in a fluid state of Y-shapes dispersed in the buffer solution. This behaviour also confirms simulation studies showing that DNA-nano-stars form thermally reversible, equilibrium hydrogels.

The thermally reversible viscoelasticity in DNA hydrogels has important consequences for studies of stability of DNA-rich networks, and may also be relevant in the development of a new class of hydrogels with more ordered structure. These could be achieved by introducing more rigid DNA-building blocks. Such ordered structures could be envisaged as builders of thermo-responsive networks loaded with drugs that could be released in a controlled fashion, or a micron-sized actuators with well-defined elastic modulus.

\section{acknowledgments}
	The authors thank D. Frenkel and C. Ness for useful discussions, P. Li for fabricating the macroscopic DNA network model, and N. Hawkins for providing technical assistance in DWS measurements. Z. X. receives financial supports from National University of Defense Technology Scholarship at Cambridge, and NanoDTC Associate Programme. E. E. and A. C. acknowledge support from the ETN-COLLDENSE (H2020-MCSA-ITN-2014, grant no. 642774). E. E. and T. W. thank the Winton Program for Sustainable Physics. T. C. and D. L. thank the National Basic Research Program of China (973 program, No. 2013CB932803), the National Natural Science Foundation of China (No. 21534007), and the Beijing Municipal Science \& Technology Commission for financial supports. I. S. and R. L. acknowledge support from EPSRC. M. Z. is funded by a joint EPSRC and Unilever CASE award RG748000.

\bibliography{arXiv_Xing}

\end{document}